\documentclass[10pt,letterpaper,twocolumn,conference]{IEEEtran}
\usepackage[cmex10]{amsmath}
\usepackage{amssymb,amsthm}
\usepackage{url}
\usepackage[dvips]{graphicx}
\usepackage{psfrag}
\usepackage{stfloats}

\newtheorem{theorem}{Theorem}
\newtheorem{lemma}{Lemma}
\newtheorem{definition}{Definition}
\newtheorem{prop}{Proposition}

\title{The Asymptotic Bit Error Probability of LDPC Codes for the Binary Erasure Channel with Finite Iteration Number}
\author{\IEEEauthorblockN{Ryuhei Mori\IEEEauthorrefmark{1}, Kenta Kasai\IEEEauthorrefmark{2}, Tomoharu Shibuya\IEEEauthorrefmark{3}, and Kohichi Sakaniwa\IEEEauthorrefmark{2}}
\IEEEauthorblockA{\IEEEauthorrefmark{1}
Dept.\ of Computer Science,
Tokyo Institute of Technology \\
Email: mori@comm.ss.titech.ac.jp
}
\IEEEauthorblockA{\IEEEauthorrefmark{2}
Dept.\ of Communications and Integrated Systems,
Tokyo Institute of Technology\\
Email: \{kenta, sakaniwa\}@comm.ss.titech.ac.jp
}
\IEEEauthorblockA{\IEEEauthorrefmark{3}
R \& D Department, National Institute of Multimedia Education\\
Email: tshibuya@nime.ac.jp
}
}

\def\Pb{\mathrm{P_b}}
\def\Pn{\mathbb{P}_n}

\begin{document}
\maketitle
%\IEEEpeerreviewmaketitle

%\begin{IEEEkeywords}
%LDPC
%\end{IEEEkeywords}
\begin{abstract}
We consider communication over the binary erasure channel (BEC) using low-density parity-check (LDPC) code and belief propagation (BP) decoding.
The bit error probability for infinite block length is known by density evolution \cite{cap} and
it is well known that a difference between the bit error probability at finite iteration number for finite block length $n$ and for infinite block length is asymptotically
$\alpha/n$, where $\alpha$ is a specific constant depending on the degree distribution, the iteration number and the erasure probability.
Our main result is to derive an efficient algorithm for calculating $\alpha$ for regular ensembles.
The approximation using $\alpha$ is accurate for $(2,r)$-regular ensembles even in small block length.
\end{abstract}

\section{Introduction}
In this paper, we consider irregular low-density parity-check (LDPC) codes \cite{gallager} with a degree distribution pair $(\lambda,\rho)$ \cite{luby}.
The bit error probability of LDPC codes over the binary erasure channel (BEC) under belief propagation (BP) decoding
is determined by three quantities;
the block length $n$, the erasure probability $\epsilon$ and the iteration number $t$.
Let $\Pb(n,\epsilon,t)$ denote the bit error probability of LDPC codes with block length $n$ over the BEC with erasure probability $\epsilon$ at iteration number $t$.
For infinite block length, $\Pb(\infty, \epsilon, t) \triangleq \lim_{n\to\infty}\Pb(n,\epsilon,t)$ can be calculated easily by density evolution \cite{cap} and
there exists threshold parameter $\epsilon_\mathrm{BP}$ such that $\lim_{t\to\infty}\Pb(\infty,\epsilon,t)=0$
for $\epsilon < \epsilon_\mathrm{BP}$ and $\lim_{t\to\infty}\Pb(\infty,\epsilon,t)>0$ for $\epsilon > \epsilon_\mathrm{BP}$.
Despite the ease of analysis for infinite block length, finite-length analysis is more complex. % and computationally intensive.
For finite block length and infinite iteration number, $\Pb(n,\epsilon,\infty) \triangleq \lim_{t\to\infty}\Pb(n,\epsilon,t)$ can be calculated exactly by \textit{stopping sets} analysis \cite{di}.
For finite block length and finite iteration number, $\Pb(n,\epsilon,t)$ can also be calculated exactly in a combinatorial way \cite{fde}.
%This expression is combinatorial therfore analytic asympototic expression is desired for optimization.
The exact finite-length analysis becomes computationally challenging as block length increasing.
An alternative approach which approximates the bit error probability is therefore employed.
For asymptotic analysis of the bit error probability, two regions of $\epsilon$ can be distinguished in the error probability;
the high error probability region called \textit{waterfall} and the low error probability region called \textit{error floor}. 
In terms of block length, they correspond to the small block length region and the large block length region.
This paper deals with the bit error probability for large block length both below and above threshold with finite iteration number.

For infinite iteration number, the asymptotic analysis for error floor was shown by Amraoui as following \cite{amraoui}:
\begin{equation*}
\Pb(n,\epsilon,\infty) = \frac1{2}\frac{\lambda'(0)\rho'(1)\epsilon}{1-\lambda'(0)\rho'(1)\epsilon}\frac1{n} + o\left(\frac1{n}\right)
\text{,\hspace{1.6em}for } \epsilon < \epsilon_\mathrm{BP} \text{,}
\end{equation*}
as $n\to\infty$.
This equation means that for ensembles with $\lambda_2 > 0$,
${\frac1{2}\frac{\lambda'(0)\rho'(1)\epsilon}{1-\lambda'(0)\rho'(1)\epsilon}\frac1{n}}$ is a good approximation of $\Pb(n,\epsilon,\infty)$
where $n$ is sufficiently large.

Our main result is following. \\
\textit{For regular LDPC codes with finite iteration number}
\begin{equation*}
\Pb(n,\epsilon,t) = \Pb(\infty,\epsilon,t) + \alpha(\epsilon,t)\frac1{n} + o\left(\frac1{n}\right) \text{,\hspace{2em}for any } \epsilon \text{,}
\end{equation*}
\textit{as $n\to\infty$, where $\alpha(\epsilon,t) = \beta(\epsilon,t) + \gamma(\epsilon,t)$ and $\beta(\epsilon,t)$ and $\gamma(\epsilon,t)$ are given by Theorem \ref{beta} and Theorem \ref{gamma}.}

This analysis is the first asymptotic analysis for finite iteration number.

\section{Main result}
\begin{figure*}[!b]
\hrulefill
%\begin{theorem}[The coefficient of $n^{-1}$ in the bit error probability due to single-cycle neighborhood graphs]\label{gamma}
%$\gamma(\epsilon,t)$ for irregular LDPC ensembles with degree distribution pair $(\lambda,\rho)$ are calculated as following
%\begin{align*}
%\gamma(\epsilon,t) &= \sum_{s_1=1}^{t-1}\sum_{s_2=2s_1+1}^{2t} F_{12}(t,s_1,s_2) +
%\sum_{s_1=0}^{t-1} \sum_{s_2=2s_1+2}^{2t} F_{34}(t,s_1,s_2) + % + F_2(t,t-1,t)
%\sum_{s=1}^{2t} F_{56}(t,s)
%\end{align*}
%where
\begin{align}
f(t,s,p) &\triangleq
\begin{cases}
\epsilon, &\text{if } t = 0\\
\epsilon \frac{\lambda'(P(t))}{\lambda'(1)} g(t,s-1,p), &\text{otherwise}
\end{cases}\text{ ,\hspace{2em}}
g(t,s,p) \triangleq
\begin{cases}
p, &\text{if } s = 0\\
1-\frac{\rho'(1-Q(t))}{\rho'(1)} (1-f(t-1,s,p)), &\text{otherwise}\\
\end{cases}\nonumber\\
%H_1(t,s) &\triangleq \epsilon\frac{\lambda''(P(t))}{\lambda''(q)}G_1(t,t,s-1)\text{ ,\hspace{6.4em}}
H(u,t,s) &\triangleq
\begin{cases}
\epsilon\frac{\lambda'(P(t))}{\lambda'(1)}G_2(t,u)\text{,}&\text{if }s=0\\
\epsilon\frac{\lambda'(P(t))}{\lambda'(1)}G_1(u,t,s-1)\text{,}&\text{otherwise}
\end{cases}\nonumber\\
G_1(u,t,s) &\triangleq \begin{cases}
G_2'(t,u),&\text{if } s = 0\\
\left(1-\frac{\rho'(1-Q(t))}{\rho'(1)}\right)g(u,u-t+s,1) + \frac{\rho'(1-Q(t))}{\rho'(1)}H(u,t-1,s-1),&\text{otherwise}\\
\end{cases}\nonumber\\
G_2(u,t) &\triangleq \begin{cases}
1,&\text{if } t= u+1\\
\left(1-\frac{\rho'(1-Q(t))}{\rho'(1)}\right)g(u,t-u-1,1) + \frac{\rho'(1-Q(t))}{\rho'(1)}\epsilon\frac{\lambda'(P(t-1)))}{\lambda'(1)}G_2(u,t-1),&\text{otherwise}\\
\end{cases}\nonumber\\
G_2'(u,t) &\triangleq \begin{cases}
1,&\text{if } t=u\\
\left(1-\frac{\rho'(1-Q(t))}{\rho'(1)}\right)g(u,t-u,1) + \frac{\rho'(1-Q(t))}{\rho'(1)}\epsilon\frac{\lambda'(P(t-1)))}{\lambda'(1)}G_2'(u,t-1),&\text{otherwise}\\
\end{cases}\nonumber\\
H_2(t,s) &\triangleq \begin{cases}
1- \frac{\rho''(1-Q(t))}{\rho''(1)}(1-\epsilon\frac{\lambda'(P(t-1))}{\lambda'(1)}),&\text{if }s=1\\
1- \frac{\rho''(1-Q(t))}{\rho''(1)}\left(1-2f(t-1,s,1)+ \epsilon\frac{\lambda'(P(t-1))}{\lambda'(1)}H(t-1,t-1,s-1)\right),&\text{otherwise}
\end{cases}\nonumber\\
F_{12}(t,s_1,s_2) &\triangleq \frac1{2}\lambda''(1)\rho'(1)^2(\lambda'(1)\rho'(1))^{s_2-s_1-2}
%\epsilon \frac{L'(P(t))}{L'(1)}\\
Q(t+1)\nonumber\\
&\quad g\left(t,s_1-1,1-\frac{\rho'(1-Q(t-s_1+1))}{\rho'(1)}\left(1-\epsilon\frac{\lambda''(P(t-s_1))}{\lambda''(1)}G_1(t-s_1,t-s_1,s_2-2s_1-1)\right)\right)\label{f12}\\
F_{34}(t,s_1,s_2) &\triangleq \frac1{2}\rho''(1)\lambda'(1)(\lambda'(1)\rho'(1))^{s_2-s_1-2}
%\epsilon \frac{L'(P(t))}{L'(1)}
Q(t+1) g\left(t,s_1, H_2(t-s_1,s_2-2s_1-1)\right)\label{f34}\\
F_{56}(t,s) &\triangleq \frac1{2}(\lambda'(1)\rho'(1))^{s}
H(t,t,s)\label{f56}
%\epsilon\frac{L''(P(t))}{L''(1)}
%\left(\left(1-\frac{\rho'(1-Q(t))}{\rho'(1)}\right) g(t,s-1,1) +
%\frac{\rho'(1-Q(t))}{\rho'(1)}h(t,t-1,s-1)\right)\\
%
%\alpha &\triangleq \sum_{s_1=1}^{t-1}\sum_{s_2=s_1+1}^t F_1(t,s_1,s_2) + \sum_{s_1=0}^{t-1} \sum_{s_2=s_1+1}^t F_2(t,s_1,s_2) +
%\sum_{s_1=1}^{t-1} \sum_{s_2=s_1}^{t-1} F_3(t,s_1,s_2) + \sum_{s_1=0}^{t-2} \sum_{s_2=s_1+1}^{t-1} F_4(t,s_1,s_2) +\\
%&\sum_{s=1}^t F_5{t,s} + \sum_{s=0}^{t-1} F_6{t,s}
%\sum_{s_1=0}^{t-1} \sum_{s_2=s_1+1}^t F_2(t,s_1,s_2) +
%\sum_{s_1=0}^{t-2} \sum_{s_2=s_1+1}^{t-1} F_4(t,s_1,s_2) +\\
%\gamma(\epsilon,t) &\triangleq \sum_{s_1=1}^{t-1}\sum_{s_2=2s_1+1}^{2t} F_{13}(t,s_1,s_2) +
%\sum_{s_1=0}^{t-1} \sum_{s_2=2s_1+2}^{2t} F_{24}(t,s_1,s_2) + % + F_2(t,t-1,t)
%\sum_{s=1}^{2t} F_{56}(t,s)
\end{align}
%\end{theorem}
%\hrulefill
\end{figure*}

The error probability of a bit in fixed tanner graph at the $t$-th iteration is determined by neighborhood graph of depth $t$ of the bit \cite{montanari,mct}.
Since the probability of neighborhood graphs which have $k$ cycles is $\Theta(n^{-k})$
we focus on the neighborhood graphs with no cycle and single cycle for calculating the coefficient of $n^{-1}$ in the bit error probability. 
Let $\beta(\epsilon,t)$ denote the coefficient of $n^{-1}$ in the bit error probability due to cycle-free neighborhood graphs and
$\gamma(\epsilon,t)$ denote the coefficient of $n^{-1}$ in the bit error probability due to single-cycle neighborhood graphs.
Then the coefficient of $n^{-1}$ in the bit error probability can be expressed as following: $\alpha(\epsilon,t)=\beta(\epsilon,t)+\gamma(\epsilon,t)$.
$\gamma(\epsilon,t)$ can be calculated efficiently for irregular ensembles and $\beta(\epsilon,t)$ can be expressed simply for regular ensembles.

\if0
Where block length tends to infinity, the neighborhood graph takes no cycle with probability $1$. In other words,
$\lim_{n\to\infty}\mathrm{P_n}(G) = 0$ for any neighborhood graph with some cycles and
\begin{equation}\label{tree}
\lim_{n\to\infty}\mathrm{P_n}(G) = L_{l_0}\prod_{i\in G\backslash 0}\lambda_{l_i}\prod_{j\in G}\rho_{r_j}
\end{equation}
for any cycle-free neighborhood graph $G$ where $l_0$ is degree of the root node
\fi
%We will call \textit{cycle free ensemble} for the ensemble of cycle free neighborhood graph with probability in (\ref{tree}).
The expected probability of erasure message for infinite block length can be calculated by density evolution.
\begin{prop}[Density evolution \cite{cap}]
Let $Q(t)$ denote erasure probability of messages into check node at the $t$-th iteration and
$P(t)$ denote erasure probability of messages into variable node at the $t$-th iteration for infinite block length.
Then
\begin{align*}
%q(x) &\triangleq \epsilon \lambda(x)\\
%p(x) &\triangleq 1-\rho(1-x)\\
%d(x) &\triangleq q(p(x))\\
Q(t) &= \epsilon\lambda(P(t-1))\\
P(t) &= 
\begin{cases}
1,&\text{if } t = 0\\
1-\rho(1-Q(t)),&\text{otherwise}\\
\end{cases}
\end{align*}
%when
%\begin{align*}
%d(x) &\triangleq \epsilon \lambda(1-\rho(1-x))\\
%\end{align*}
\end{prop}

\if0
\begin{definition}[The probability]
\begin{align*}
%f_1(t,s,x) &\triangleq \epsilon \frac{\lambda'(P(t))}{\lambda'(1)} (1-g_1(t,s,x))\\
%g_1(t,s,x) &\triangleq \frac{\rho'(1-Q(t))}{\rho'(1)} (1-f_1(t-1,s-1,x))\\
%f_1(t,0,x) &\triangleq x\\
%g_1(t,0,x) &\triangleq 0\\
f(t,s,p) &\triangleq \epsilon \frac{\lambda'(P(t))}{\lambda'(1)} g(t,s-1,p)\\
f(0,s,p) &\triangleq \epsilon\\
g(t,s,p) &\triangleq 1-\frac{\rho'(1-Q(t))}{\rho'(1)} (1-f(t-1,s,p))\\
%f(t,0,p) &\triangleq 0\\
g(t,0,p) &\triangleq p\\
%\\
%f_3(t,s,x) &\triangleq \epsilon \frac{\lambda'(P(t))}{\lambda'(1)} (1-g_3(t,s-1,x))\\
%g_3(t,s,x) &\triangleq \frac{\rho'(1-Q(t))}{\rho'(1)} (1-f_3(t-1,s-1,x))\\
%f_3(t,0,x) &\triangleq x\\
%g_3(t,0,x) &\triangleq x\\
\end{align*}
\end{definition}
\fi

%\onecolumn
%\begin{figure*}[!h]
%\normalsize
%test
%\hrulefill
%\end{figure*}
\begin{figure*}[!t]
\psfrag{marginalized}{marginalized}
\psfrag{l1}{$l_1$}
\psfrag{l2}{$l_2$}
\psfrag{l3}{$l_3$}
\psfrag{l4}{$l_4$}
\psfrag{l5}{$l_5$}
\psfrag{l6}{$l_6$}
\psfrag{l7}{$l_7$}
\psfrag{l8}{$l_8$}
\includegraphics[height=15em]{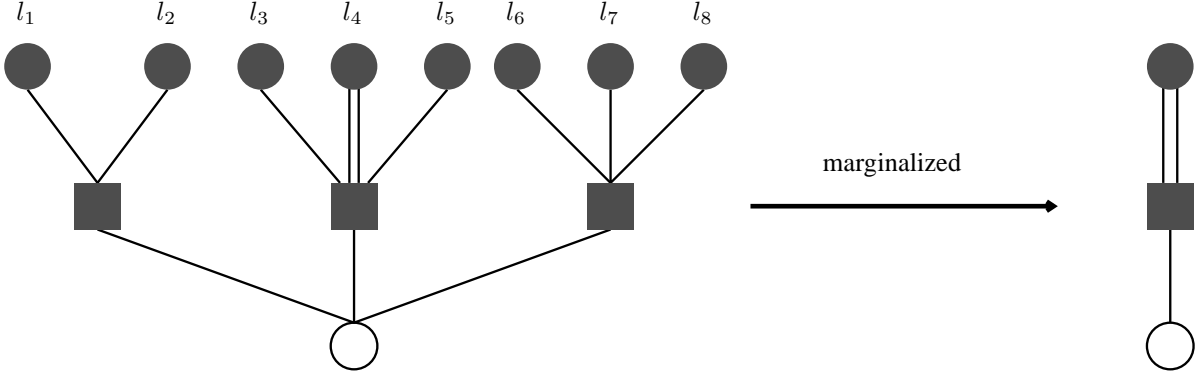}
\caption{The left figure is the neighborhood graph of depth 1. White variable node is the root node. The variable nodes in depth 1 have degree $l_1$ to $l_8$.
We consider summing up the probability of all nodes and the type of marginalized graph in the right figure.}
\label{exm}
\end{figure*}

%$\gamma(\epsilon,t)$ can be calculated by Theorem \ref{gamma}. Outline of the proof is in section \ref{th1}.
The coefficient of $n^{-1}$ in the bit error probability due to single-cycle neighborhood graphs can be calculated using density evolution.% as Theorem \ref{gamma}.
\begin{theorem}[The coefficient of $n^{-1}$ in the bit error probability due to single-cycle neighborhood graphs]\label{gamma}
$\gamma(\epsilon,t)$ for irregular LDPC ensembles with a degree distribution pair $(\lambda,\rho)$ are calculated as following
\begin{multline*}
\gamma(\epsilon,t) = \sum_{s_1=1}^{t-1}\sum_{s_2=2s_1+1}^{2t} F_{12}(t,s_1,s_2) +\\
\sum_{s_1=0}^{t-1} \sum_{s_2=2s_1+2}^{2t} F_{34}(t,s_1,s_2) + % + F_2(t,t-1,t)
\sum_{s=1}^{2t} F_{56}(t,s)
\end{multline*}
where $F_{12}$, $F_{34}$ and $F_{56}$ is Eq. (\ref{f12}), (\ref{f34}) and (\ref{f56}), respectively.
\end{theorem}
The complexity of the computation of $\gamma(\epsilon,t)$ is $O(t^3)$ in time and $O(t^2)$ in space.

$\beta(\epsilon,t)$ can be expressed simply for regular ensembles since of uniqueness of the cycle-free neighborhood graph.
\begin{theorem}[The coefficient of $n^{-1}$ in the bit error probability due to cycle-free neighborhood graphs for regular ensembles]\label{beta}
$\beta(\epsilon,t)$ for the $(l,r)$-regular LDPC ensemble is expressed as following
\begin{multline*}
\beta(\epsilon,t) = -\frac1{2}l(r-1)\frac{1-\{(l-1)(r-1)\}^t}{1-(l-1)(r-1)}\{(l-1)(r-1)\}^t\\ \epsilon P(t)^l\text{.}
\end{multline*}
\end{theorem}
\begin{IEEEproof}[Outline of the proof]
The probability of the unique cycle-free neighborhood graph of depth $t$ is
\begin{equation*}
\frac{\prod_{i=0}^{l\frac{1-\{(l-1)(r-1)\}^t}{1-(l-1)(r-1)}-1} (E-ir)\prod_{i=1}^{l(r-1)\frac{1-\{(l-1)(r-1)\}^t}{1-(l-1)(r-1)}}(E-il)}
{\prod_{i=0}^{lr\frac{1-\{(l-1)(r-1)\}^t}{1-(l-1)(r-1)}-1}(E-i)}\text{,}
\end{equation*}
where $E\triangleq nl$.
The coefficient of $n^{-1}$ in the probability is
\begin{equation*}
-\frac1{2}l(r-1)\frac{1-\{(l-1)(r-1)\}^t}{1-(l-1)(r-1)}\{(l-1)(r-1)\}^t
\end{equation*}
and the error probability of the root node is $\epsilon P(t)^l$.
Then we obtain the statement of the theorem.
\end{IEEEproof}

Due to the above theorems, $\alpha(\epsilon,t)$ for regular ensembles can be calculated efficiently.

\begin{prop}[The bit error probability decays exponentially \cite{mct}]
Assume $\epsilon < \epsilon_\mathrm{BP}$. Then for any $\delta > 0$, there exists some iteration number $T > 0$ such that for any $t \ge T$
\begin{align*}
P(t) &\le P(T) (\lambda'(0)\rho'(1)\epsilon + \delta)^{t-T}\\
P(t) &\ge P(T) (\lambda'(0)\rho'(1)\epsilon - \delta)^{t-T}\text{.}
\end{align*}
\end{prop}
Although if $\lambda'(0)\lambda'(1)\rho'(1)^2\epsilon < 1$ then $\beta(\epsilon,t)$ converges to $0$ and
$\gamma(\epsilon,t)$ converges to $\frac1{2}\frac{\lambda'(0)\rho'(1)\epsilon}{1-\lambda'(0)\rho'(1)\epsilon}$ as $t\to\infty$,
if $\lambda'(0)\lambda'(1)\rho'(1)^2\epsilon > 1$ then $\beta(\epsilon,t)$ and $\gamma(\epsilon,t)$ grow exponentially as $t\to\infty$ due to the above proposition.
Thus convergence of $\alpha(\epsilon,t)$ is non-trivial.
In practice it is necessary to use high precision floating point tools for calculating $\alpha(\epsilon,t)$.
%In the numerical calculation in the next secion, we use GNU MP with the precision 1024.

%\twocolumn
%\newpage
%\onecolumn
%\clearpage
%\begin{multicols}{2}
\section{Outline of the proof of theorem \ref{gamma}}\label{th1}
\begin{figure*}[!t]
\psfrag{t1}{$s_1=1$, $s_2=6$}
\psfrag{t2}{$s_1=1$, $s_2=5$}
\psfrag{t3}{$s_1=0$, $s_2=4$}
\psfrag{t4}{$s_1=0$, $s_2=5$}
\psfrag{t5}{$s=6$}
\psfrag{t6}{$s=5$}
\includegraphics[width=\hsize]{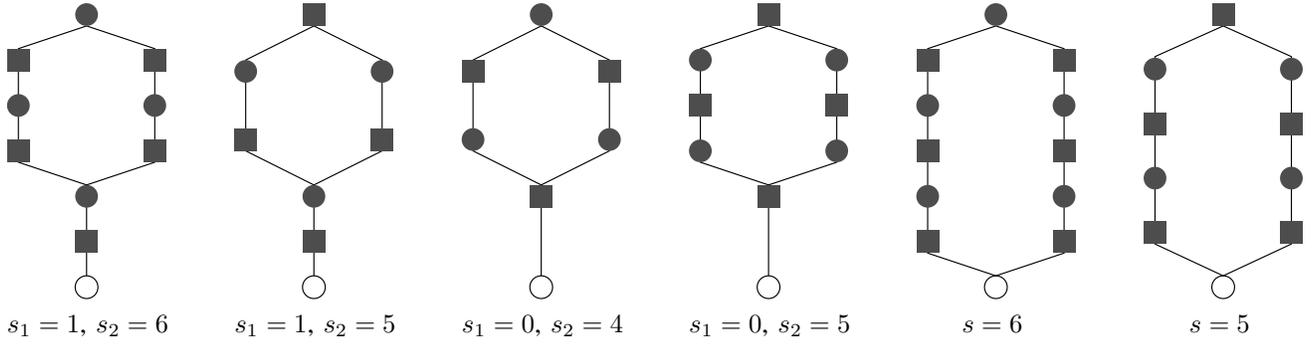}
\caption{Six types of marginalized single-cycle neighborhood graphs.
These are distinguished in which variable node, check node or root node are bifurcation node and which variable node or check node are confluence node.
Depth of the bifurcation node corresponds to $s_1$.
The number of nodes in the minimum path from root node to confluence node corresponds to $s_2$ and $s$.}
\label{sc}
\end{figure*}

The bit error probability of an ensemble with iteration number $t$ is defined as following:
\begin{equation*}
\Pb(n,\epsilon,t) \triangleq \sum_{G\in\mathcal{G}_t}\Pn(G)\Pb(G)\text{,}
\end{equation*}
where $\mathcal{G}_t$ denotes a set of all neighborhood graphs of depth $t$, $\Pn(G)$ denotes the probability of the neighborhood graph $G$
and $\Pb(G)$ denotes the error probability of the root node in the neighborhood graph $G$.
The coefficient of $n^{-1}$ in the bit error probability with iteration number $t$ due to single-cycle neighborhood graphs is defined as following:
\begin{equation*}
\gamma(\epsilon,t) \triangleq \lim_{n\to\infty}n\sum_{G\in\mathcal{S}_t}\Pn(G)\Pb(G)\text{,}
\end{equation*}
where $\mathcal{S}_t$ denotes a set of all single-cycle neighborhood graphs of depth $t$.

First we consider the bit error probability of the root node of the neighborhood graph $G$ in Fig. \ref{exm}.
The variable nodes in depth 1 have degree $l_1$ to $l_8$.
%Let $\Pn(G)$ denote the probability of the neighborhood graph $G$. Then the coefficient of $n^{-1}$ in $\Pn(G)$ is given as
Then the coefficient of $n^{-1}$ in $\Pn(G)$ is given as
\begin{multline*}
\lim_{n\to\infty} n\Pn(G) = \frac1{L'(1)}L_3\rho_3\rho_5\rho_4\\
\lambda_{l_1}\lambda_{l_2}\lambda_{l_3}\lambda_{l_4}\lambda_{l_5}\lambda_{l_6}\lambda_{l_7}\lambda_{l_8}(l_4-1) \text{.}
\end{multline*}
The error probability of the message from the channel to the root node is $\epsilon$.
The error probabilities of the message from the left check node, the right check node and the middle check node to the root node are
$(1-(1-\epsilon)^2)$, $(1-(1-\epsilon)^3)$ and $(1-(1-\epsilon)^3)$, respectively.
Then the error probability of the root node is given as
\begin{equation*}
\mathrm{P_b}(G) = \epsilon(1-(1-\epsilon)^2)(1-(1-\epsilon)^3)(1-(1-\epsilon)^3) \text{.}
%\epsilon(1-(1-\epsilon)^2)(1-(1-\epsilon)^3)(1-(1-\epsilon)^3).
\end{equation*}
The coefficient of $n^{-1}$ term of the bit error probability due to $G$ is given as
\begin{multline*}
\lim_{n\to\infty}n\Pn(G)\Pb(G) = \\
\frac1{L'(1)}L_3\rho_3\rho_5\rho_4 \lambda_{l_1}\lambda_{l_2}\lambda_{l_3}\lambda_{l_4}\lambda_{l_5}\lambda_{l_6}\lambda_{l_7}\lambda_{l_8}(l_4-1)\\
\epsilon(1-(1-\epsilon)^2)(1-(1-\epsilon)^3)(1-(1-\epsilon)^3) \text{.}
\end{multline*}
After summing out the left and right subgraphs,
\begin{equation*}
\frac1{L'(1)}L_{3}\rho_5\lambda_{l_3}\lambda_{l_4}\lambda_{l_5}(l_4-1)
\epsilon(1-(1-\epsilon)^3)P(1)^2.
\end{equation*}
After summing out degrees $l_3$, $l_4$ and $l_5$,
\begin{equation*}
\frac1{L'(1)}L_{3}\rho_5\lambda'(1) \epsilon(1-(1-\epsilon)(1-Q(1))^2)P(1)^2 \text{.}
\end{equation*}
%
%Marginal probability of the left and the right check nodes is 
%\begin{align*}
%&\frac{\lambda'(1)}{L'(1)}L_{3}\rho_4 \epsilon(1-(1-\epsilon)^2)\left(\sum_r \rho_r(1-(1-\epsilon)^{r-1})\right)^2\\
%&= \frac{\lambda'(1)}{L'(1)}L_3\rho_4\epsilon(1-(1-\epsilon)^2) (1-\rho(1-\epsilon))^2\\
%&= \frac{\lambda'(1)}{L'(1)}L_3\rho_4\epsilon(1-(1-\epsilon)^2) P(1)^2
%\end{align*}
%
At last, after summing out the root node and the middle check node,
\begin{align*}
&\sum_{l,r}\frac{\lambda'(1)}{L'(1)}L_l\rho_r\epsilon(1-(1-Q(1))^{r-3}(1-\epsilon))\\
&\quad P(1)^{l-1}l\binom{r-1}{2}\\
& = \frac{\lambda'(1)}{2L'(1)}L'(P(1))\epsilon(\rho''(1)-\rho''(1-Q(1))(1-\epsilon))\\
%& = \frac{\lambda'(1)}{2L'(1)}L'(P(1))\epsilon(\rho''(1)-\rho''(1-Q(1))(1-\epsilon))\\
& = \frac1{2}\lambda'(1)\rho''(1)
\epsilon\frac{L'(P(1))}{L'(1)} \left(1-\frac{\rho''(1-Q(1))}{\rho''(1)}\left(1-\epsilon\right)\right).
\end{align*}
The coefficient of $n^{-1}$ in the bit error probability for iteration number $t$ due to neighborhood graphs with the right graph type in Fig. \ref{exm} is given as
\begin{align*}
&\frac1{2}\lambda'(1)\rho''(1)\\
&\quad \epsilon\frac{L'(P(t))}{L'(1)}\left(1-\frac{\rho''(1-Q(t))}{\rho''(1)}\left(1-\epsilon\frac{\lambda'(P(t-1))}{\lambda'(1)}\right)\right)\\
&=\frac1{2}\lambda'(1)\rho''(1)Q(t+1)g(t, 0, H_2(t,1))\\
&=F_{34}(t,0,2)
\end{align*}
in the same way.
Notice that $\frac1{2}\lambda'(1)\rho''(1)$ is the coefficient of $n^{-1}$ of the probability of neighborhood graphs with the right graph type in Fig. \ref{exm}.

Single-cycle neighborhood graphs can be classified to six types in Fig. \ref{sc}.
Summing up the bit error probability due to all these types, we obtain $\gamma(\epsilon,t)$.
Left two types correspond to $F_{12}$, middle two types correspond to $F_{34}$ and right two types correspond to $F_{56}$.

\section{Numerical calculations and simulations}
\begin{figure}[h]
\psfrag{epsilon}{$\epsilon$}
\includegraphics[width=\hsize]{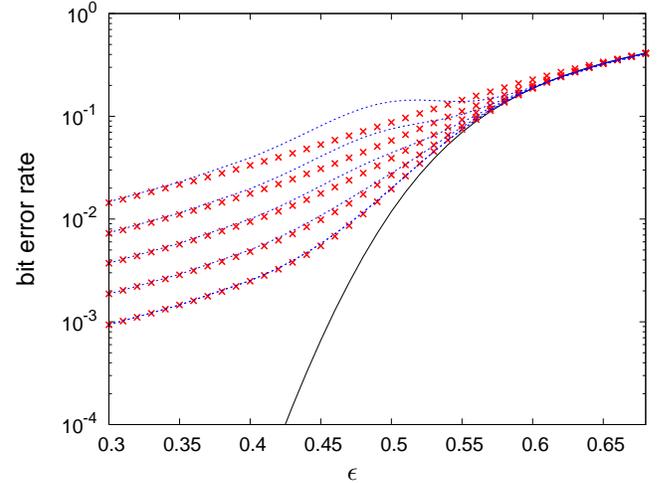}
\caption{Comparing $\Pb(\infty,\epsilon,t) + \alpha(\epsilon,t)\frac1{n}$ with numerical simulations for the $(2,3)$-regular ensemble with iteration number 20.
The dotted curves are approximation and the solid curve is density evolution. Block lengths are 51, 102, 201, 402 and 801. The threshold is 0.5.}
\label{reg23}
\end{figure}

There is a question that \textit{how large block length is necessary for using $\Pb(\infty,\epsilon,t) + \alpha(\epsilon,t)\frac1{n}$ for a good approximation of $\Pb(n,\epsilon,t)$}.
It is therefore interesting to compare $\Pb(\infty,\epsilon,t) + \alpha(\epsilon,t)\frac1{n}$ with numerical simulations.
In the proof, we count only the error probability due to cycle-free neighborhood graphs and single-cycle neighborhood graphs.
Thus it is expected that the approximation is accurate only at large block length
where the probability of the multicycle neighborhood graphs is sufficiently small.
Contrary to the expectation, the approximation is accurate already at small block length in Fig. \ref{reg23}.
Although there is a large difference in small block length near the threshold, the approximation is accurate at block length 801 which is not large enough.

\begin{figure}[ht]
\psfrag{epsilon}{$\epsilon$}
\includegraphics[width=\hsize]{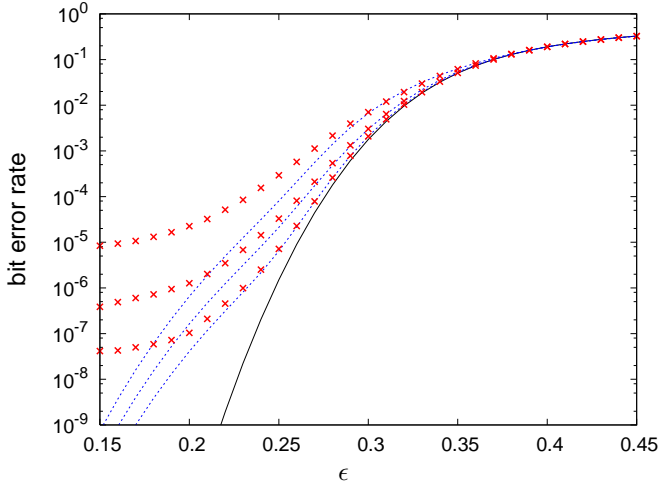}
\caption{Comparing $\Pb(\infty,\epsilon,t) + \alpha(\epsilon,t)\frac1{n}$ with numerical simulations for the $(3,6)$-regular ensemble with iteration number 5.
The dotted curves are approximation and the solid curve is density evolution. Block lengths are 512, 2048 and 8192. The threshold is 0.42944.}
\label{reg36}
\end{figure}

For the ensembles with $\lambda_2 = 0$, the approximation is not accurate at $\epsilon$ far below the threshold in Fig. \ref{reg36}.
Since $|\alpha(\epsilon,t)|$ decreases to $0$ as $t\to\infty$ for the ensembles
the higher order terms caused by multicycle stopping sets has a large contribution to the bit error probability.
%the $n^{-2}$ term which is caused by three double cycles stopping sets has large effect on the bit error probability for the $(3,6)$-regular case.
%It is expected that the approximation is even accurate for ensembles with $\lambda_2=0$ from which stopping sets with small number of cycles are expurgated.
It is expected that the approximation is even accurate for the ensembles from which stopping sets with small number of cycles are expurgated.

\begin{figure}[t]
\psfrag{epsilon}{$\epsilon$}
\psfrag{coef}{the coefficient of $n^{-1}$}
\includegraphics[width=\hsize]{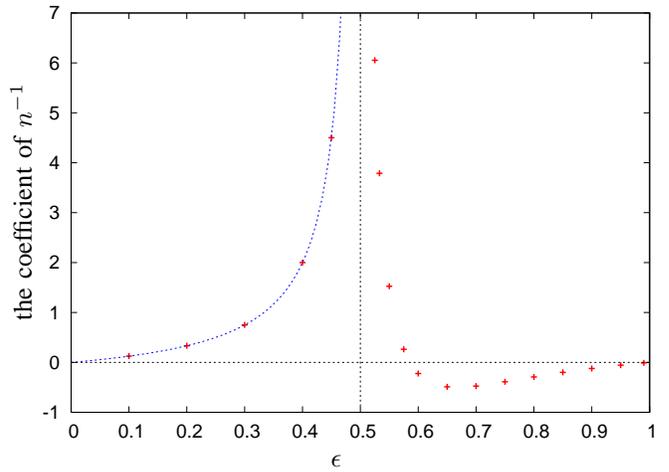}
\caption{Comparing $\alpha(\epsilon,\infty)$ with $\frac1{2}\frac{\lambda'(0)\rho'(1)\epsilon}{1-\lambda'(0)\rho'(1)\epsilon}$ for the $(2,3)$-regular ensemble.
Below the threshold 0.5, they take almost the same value.}
\label{lim23}
\end{figure}

\begin{figure}[t]
\psfrag{epsilon}{$\epsilon$}
\psfrag{coef}{the coefficient of $n^{-1}$}
\includegraphics[width=\hsize]{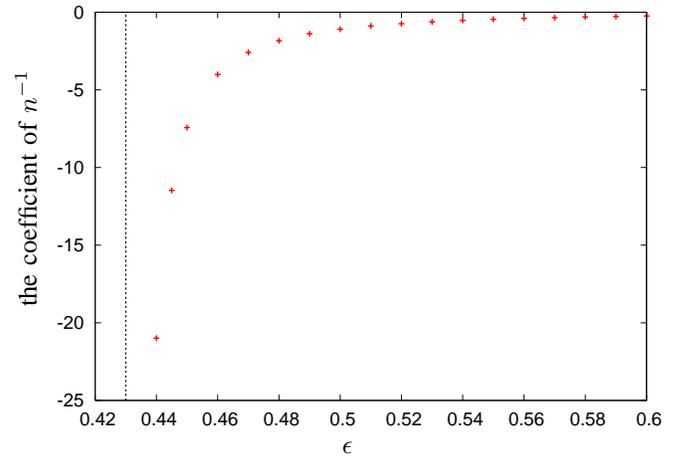}
\caption{$\alpha(\epsilon,\infty)$ plotted for the $(3,6)$-regular ensemble above the threshold 0.42944.}
\label{lim36}
\end{figure}

The limiting value of $\alpha(\epsilon,t)$, $\alpha(\epsilon,\infty)\triangleq\lim_{t\to\infty}\alpha(\epsilon,t)$ is also interesting.
For $\alpha(\epsilon,\infty)$, calculate $\alpha(\epsilon,t)$ where sufficiently large $t$ in Fig. \ref{lim23} and Fig. \ref{lim36}.
For the $(2,3)$-regular ensemble below the threshold,
$\alpha(\epsilon,\infty)$ and $\frac1{2}\frac{\lambda'(0)\rho'(1)\epsilon}{1-\lambda'(0)\rho'(1)\epsilon}$ take almost the same value.
It implies that below threshold $n(\Pb(n,\epsilon,t)-\Pb(\infty,\epsilon,t))$ % = \lim_{n\to\infty}\lim_{t\to\infty}n\Pb(n,\epsilon,t).
takes the same value at two limits; $n\to\infty$ then $t\to\infty$ and $t\to\infty$ then $n\to\infty$.
For the ensembles with $\lambda_2=0$, $\alpha(\epsilon,\infty)$ is almost $0$ where $\epsilon$ is smaller than threshold.

At last, notice that $\alpha(\epsilon,t)$ takes non-trivial values slightly below threshold.
For the $(3,6)$-regular ensemble, $\alpha(0.425,t)$ is negative at $t\le 39$, positive at $40\le t \le 52$ and has absolute value which is too small to be measured at $t = 53$.
$\max_{1\le t \le 53} |\alpha(0.425,t)| = 35710.34$ at $t = 42$.

%\onecolumn

%\begin{multicols}{2}
\section{Outlook}
Although the asymptotic analysis of the bit error probability for finite block length and finite iteration number given in this paper is very accurate at $(2,r)$-regular,
much work remains to be done.
First there remains the problem to computing $\beta(\epsilon,t)$ for irregular ensembles.
It would also be interesting to generalize this algorithm to other ensembles and other channels.

In the binary memoryless symmetric channel (BMS) parametrized by $\epsilon \in [0,\infty)$,
We consider $\inf_t \Pb(n,\epsilon,t)$ instead of $\lim_{t\to\infty} \Pb(n,\epsilon,t)$ since of lack of monotonicity.
The asymptotic analysis of the bit error probability with the best iteration number
$t^*(n,\epsilon) \triangleq \arg\inf_t \Pb(n,\epsilon,t)$ under BP decoding was shown by Montanari for small $\epsilon$ as following \cite{montanari}:
\begin{equation*}
\Pb(n,\epsilon,t^*(n,\epsilon)) = \frac1{2}\sum_{i=0}^\infty(\lambda'(0)\rho'(1))^i\mathfrak{p_i}\frac1{n} + o\left(\frac1{n}\right)
\end{equation*}
%\newpage \noindent
as $n\to\infty$, where $\mathfrak{p_i}\triangleq\Pr(Z_i<0)+\frac1{2}\Pr(Z_i=0)$ and
$Z_i$ is a random variable corresponding to the sum of the $i$ i.i.d. channel log-likelihood ratio.
It implies that if $\lambda_2>0$, the asymptotic bit error probability under BP decoding is equal to that of maximum likelihood (ML) decoding.
Although the condition of the proof in \cite{montanari} implies the convergence of values corresponding to $\beta(\epsilon,t)$ and $\gamma(\epsilon,t)$ in this paper,
in general if $\lambda'(0)\lambda'(1)\rho'(1)^2\mathfrak{B}(\epsilon)>1$, they do not converge, where $\mathfrak{B}(\epsilon)$
is Bhattacharyya constant. Although the condition of $\epsilon$ is strong, the approximation is very accurate for all $\epsilon$ smaller than threshold.
%we have problems how about finite iteration number, higher order terms, the block error probability and other ensembles for the BMS.
We have the problem to prove the convergence of $\alpha(\epsilon,t)$ for the BEC and the BMS for any $\epsilon<\epsilon_\mathrm{BP}$.

A iteration number is also important.
The approximation is not accurate for too large iteration number.
A sufficient (and necessary) iteration number for a given block length and a ensemble is very important to improve the analysis in this paper.

%\newpage

%\end{multicols}

\end{document}